\documentclass[aps,pre,nofootinbib,showpacs,twocolumn]{revtex4-1}
\usepackage{amssymb,latexsym,mathrsfs}
\usepackage{amsfonts}
\usepackage{mathrsfs}
\usepackage{amsmath}
\usepackage{graphicx}
\usepackage{color}
\usepackage{verbatim}
\usepackage{cancel}

\newcommand{\beq}{\begin{equation}}
\newcommand{\eeq}{\end{equation}}
\newcommand{\beqn}{\begin{eqnarray}}
\newcommand{\eeqn}{\end{eqnarray}}
\newcommand{\bearr}{\begin{array}}
\newcommand{\enarr}{\end{array}}

\def\bea{\begin{eqnarray}}
\def\eea{\end{eqnarray}}
\def\ba{\begin{array}}
\def\ea{\end{array}}

\begin{document}
\title{Scaling properties of $d$-dimensional complex networks}
\author{Samura\'i Brito$^1$}
\email[E-mail address: ]{samuraigab@gmail.com.br}
\author{Thiago C. Nunes$^2$}
\email[E-mail address: ]{thiago.cris@yahoo.com.br}
\author{Luciano R. da Silva$^{2,3}$}
\email[E-mail address: ]{luciano@fisica.ufrn.br}
\author{Constantino Tsallis$^{3,4,5,6}$}
\email[E-mail address: ]{tsallis@cbpf.br}
\affiliation{$^1$ International Institute of Physics, Universidade Federal do Rio Grande do Norte, Campus Universit\'ario, Lagoa Nova, Natal-RN 59078-970, Brazil}
\affiliation{$^2$Departamento de F\'{\i}sica Te\'orica e Experimental, Universidade Federal do Rio Grande do Norte, Natal, RN, 59078-900, Brazil}
\affiliation{$^3$National Institute of Science and Technology of Complex Systems, Brazil}
\affiliation{$^4$Centro Brasileiro de Pesquisas F\'isicas, Rua Xavier Sigaud 150, 22290-180 Rio de Janeiro-RJ,  Brazil}
\affiliation{$^5$ Santa Fe Institute, 1399 Hyde Park Road, New Mexico 87501, USA}
\affiliation{$^6$ Complexity Science Hub Vienna, Josefstaedter Strasse 39, A 1080 Vienna, Austria.}


\begin{abstract}
The area of networks is very interdisciplinary and exhibits many applications in several fields of science. Nevertheless, there are few studies focusing on geographically located $d$-dimensional networks. In this paper, we study scaling properties of a wide class of $d$-dimensional geographically located networks which grow with preferential attachment involving Euclidean distances through $r_{ij}^ {-\alpha_A} \;(\alpha_A \geq 0)$. We have numerically analyzed the time evolution of the connectivity of sites, the average shortest path, the degree distribution entropy, and the average clustering coefficient, for $d=1,2,3,4$,  and typical values of $\alpha_A$. Remarkably enough,  virtually all the curves can be made to collapse as functions of the scaled variable $\alpha_A/d$. These observations confirm the existence of three regimes. The first one occurs in the interval $\alpha_A/d \in [0,1]$; it is non-Boltzmannian with very-long-range interactions in the sense that the degree distribution is a $q$-exponential with $q$ constant and above unity. The critical value $\alpha_A/d =1$ that emerges in many of these properties is replaced by $\alpha_A/d =1/2$ for the $\beta$-exponent which characterizes the time evolution of the connectivity of sites. 
The second regime is still non-Boltzmannian, now with moderately long-range interactions, and reflects in an index $q$ monotonically decreasing with $\alpha_A/d$ increasing from its critical value to a characteristic value $\alpha_A/d \simeq 5$. Finally, the third regime is Boltzmannian-like (with $q \simeq 1$), and corresponds to short-range interactions.

\end{abstract}

\maketitle


\section{Introduction}

\quad\par Networks are everywhere, from the Internet to social networks. We are living in the network age and the emergence of more and more related researches is natural. The theory of networks has applications in a diversity of scientific fields, such as medicine \cite{Barabasi2011}, cosmology \cite{Boguna2014}, quantum information theory \cite{Perseguers2010} and social networks \cite{Ferreira2018}. For a long time, it was diffusely believed that the statistics governing complex networks was only the Boltzmann-Gibbs (BG) one. However, in 2005 the connection between networks and nonextensive statistical mechanics started to be explored~\cite{SoaresTsallisMarizSilva2005, ThurnerTsallis2005,Thurner2005}, and is presently very active \cite{Andradeetal2005,Lindetal2007,Mendesetal2012,Almeidaetal2013,Luciano, BritoSilvaTsallis2016, NunesBritoSilvaTsallis2017}. 

In the literature, systems with long-range interactions are characterized by paired potentials that decay slowly with the distance. A potential of the form $ 1/r^\alpha$ is typically said to be long-range if $0 \leq \alpha \leq d$, where $d$ is the spatial dimension of the system. Some examples of such potentials are gravitational systems, two-dimensional hydrodynamic systems, two-dimensional elastic systems, charged systems and dipole systems. Unlike the case of classical systems with short-range interactions (usually described within BG statistics), where many results are well understood, there is a lack of complete knowledge about the dynamic and statistical properties of systems with long-range interactions (for which BG statistics fails). In this sense, many theories have been proposed to understand the systems that interact at long-range, and $q$-statistics has shown satisfactory results for this regime \cite{TirnakliBorges2016,CirtoAssisTsallis2014,CirtoRodriguezNobreTsallis2018,CirtoLimaNobre2015,christo1,christo2}.

In 2016, we studied a $d$-dimensional network model where the interactions are short- or long-ranged depending on the choice of the parameter $\alpha_A \ge 0$. The results that were obtained reinforced the connection between nonextensive statistical mechanics and the networks theory \cite{BritoSilvaTsallis2016}. In that work, we found some quantities which present a universal behaviour with respect to the particular variable $\alpha_A/d$ and observed the existence of three regimes. In the first one, namely $0 \le \alpha_A/d \le 1$, $q$ is constant and larger than unity, characterizing a non-Boltzmannian regime with very-long-range interactions. In the second one, $q$ monotonically decreases as $\alpha_A/d$ increased from its critical value $\alpha_A/d=1$ to a characteristic value $\alpha_A/d \simeq 5$. The third regime, above this characteristic value~\cite{note1}, is Boltzmannian-like $(q\simeq 1)$ and corresponds to short-range interactions. For the $\beta$ exponent (defined here below) the behaviour is somewhat different: a first regime is exhibited for $0 \leq \alpha_A/d \leq 1/2$, a second regime appears between $ \alpha_A/d=1/2$ and a characteristic value once again close to 5, and a third regime, Boltzmannian-like with $q \simeq 1$, between this value and infinity; it cannot be excluded that the purely Boltzmannian behaviour only occurs for  $\alpha_A/d \to \infty$. 

Our model was constructed through two stages: the number of the sites increases at time and the connections between the sites follow a preferential attachment rule, given by:
\begin{equation}
\Pi_i \propto k_i{r_i}^{-\alpha_A}.
\end{equation}
Each newly created site can connect to $m$ others. In the present work, all results were obtained for $m=1$. The growth of the network starts with one site at the origin, and then, we stochastically locate a second site (and then a third, a fourth, and so on up to N) through the $d$-dimensional isotropic distribution
\begin{equation}
p(r) \propto \frac{1}{r^{d+\alpha_G}}\; (\alpha_G >0;\; d=1,2,3,4),
\end{equation}
where $r\geq1$ is the Euclidean distance from the newly arrived site to the center of mass of the pre-existing system. For more details see \cite{BritoSilvaTsallis2016}.
This network is characterized by three parameters $\alpha_A$, $\alpha_G$ and $d$, where $\alpha_A$ controls the importance of the distance in the preferential attachment rule, $\alpha_G$ is associated with the geographical distribution of the sites, and $d$ is the dimension of the system. 

The connectivity distribution was the only property studied in the previous work. Our results showed that the degree distribution of this model is very well described by the $q$-exponential functions that emerges from nonextensive statistical mechanics~\cite{Tsallis1988,Tsallis2009, TsallisCirto2013}, more precisely $P(k) \sim {e_q}^{-k/\kappa}$ $\forall \; (\alpha_A,\alpha_G, d)$, with $e_q^{z} \equiv [1 + (1 - q)z]^{\frac{1}{1-q}}$. The relation between $q$ and $\gamma$ (the exponent of the asymptotic power law) is given by $\gamma \equiv 1/(q-1)$ (see \cite{SoaresTsallisMarizSilva2005} for more details). When $\alpha_A = 0$ we recover the Barab\'asi-Albert (BA) model \cite{BarabasiAlbert1999} with $q = 4/3$ ($\gamma = 3$). Remarkably enough, our previous results showed that the index $q$ and $\kappa$ exhibit universal behaviours with respect to the scaled variable $\alpha_A/d$ $ (\forall d)$.

Motivated by the results in \cite{BritoSilvaTsallis2016}, in the present work we are interested in investigating, for the same network model, other possible universal behaviors with respect to the same scaled variable $\alpha_A/d$. Besides that, we also are interested in verifying the existence of the same three regimes that we have previously observed. We have analyzed the exponent $\beta$, which is associated with the time evolution of the connectivity of sites, the average shortest path $\langle l \rangle$, the degree distribution entropy $S_q$ and the average clustering coefficient $\langle C \rangle$.  Along the lines of \cite{BritoSilvaTsallis2016}, in order to analyse these properties, we choose the typical value $\alpha_G=2$ and vary the parameters $(\alpha_A, d)$.



\section{Results}
\subsection{Time evolution of the connectivity of sites}
\begin{figure}[!htb]
\centering
\includegraphics[scale=.34]{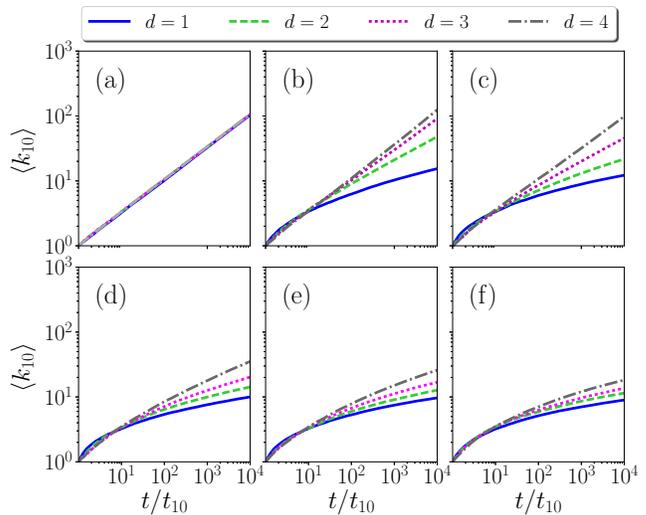}
\caption{Time evolution of the connectivity of the site $i=10$ in $\log-\log$ plot for different values of $\alpha_A$ and $d = 1,2,3,4$. The figure sublabels refer to {\bf(a)} $\alpha_A=0$, {\bf(b)} $\alpha_A=2$, {\bf(c)} $\alpha_A=3$, {\bf(d)} $\alpha_A=5$, {\bf(e)} $\alpha_A=6$ and {\bf(f)} $\alpha_A=8$. We can see that $k_i \propto (t/t_i)^{\beta(\alpha_A,d)}$ where $\beta(\alpha_A,d)$ is the asymptotic slope of the curves. For $\alpha_A = 0$, independent of the dimension, we recover the BA model with $\beta = 1/2$, and when $\alpha_A \to \infty$ the dimension does not matter either.}
\label{temp_evol}
\end{figure}

One of the most common analyses that are done in networks theory is to verify how the degree of a site changes at time. This property is usually referred to as {\it connectivity time evolution} and it usually follows the behaviour
\begin{equation}
k_i(t)\propto \left(\frac{t}{t_i}\right)^{\beta} \;\;\; \text{where}\;\; t_i \leq t.
\end{equation}
We analyzed the time evolution of the connectivity of sites for typical values of $\alpha_A$ and $d=1,2,3,4$ (see Fig.~\ref{temp_evol}). In order to do that, we choose the site $i=10$ (the result is independent of $i$), and then we compute the time evolution of its connectivity.  All simulations were made for $10^5$ sites and $10^3$ samples. 

\begin{figure}[!htb]
\centering
\includegraphics[scale=.28]{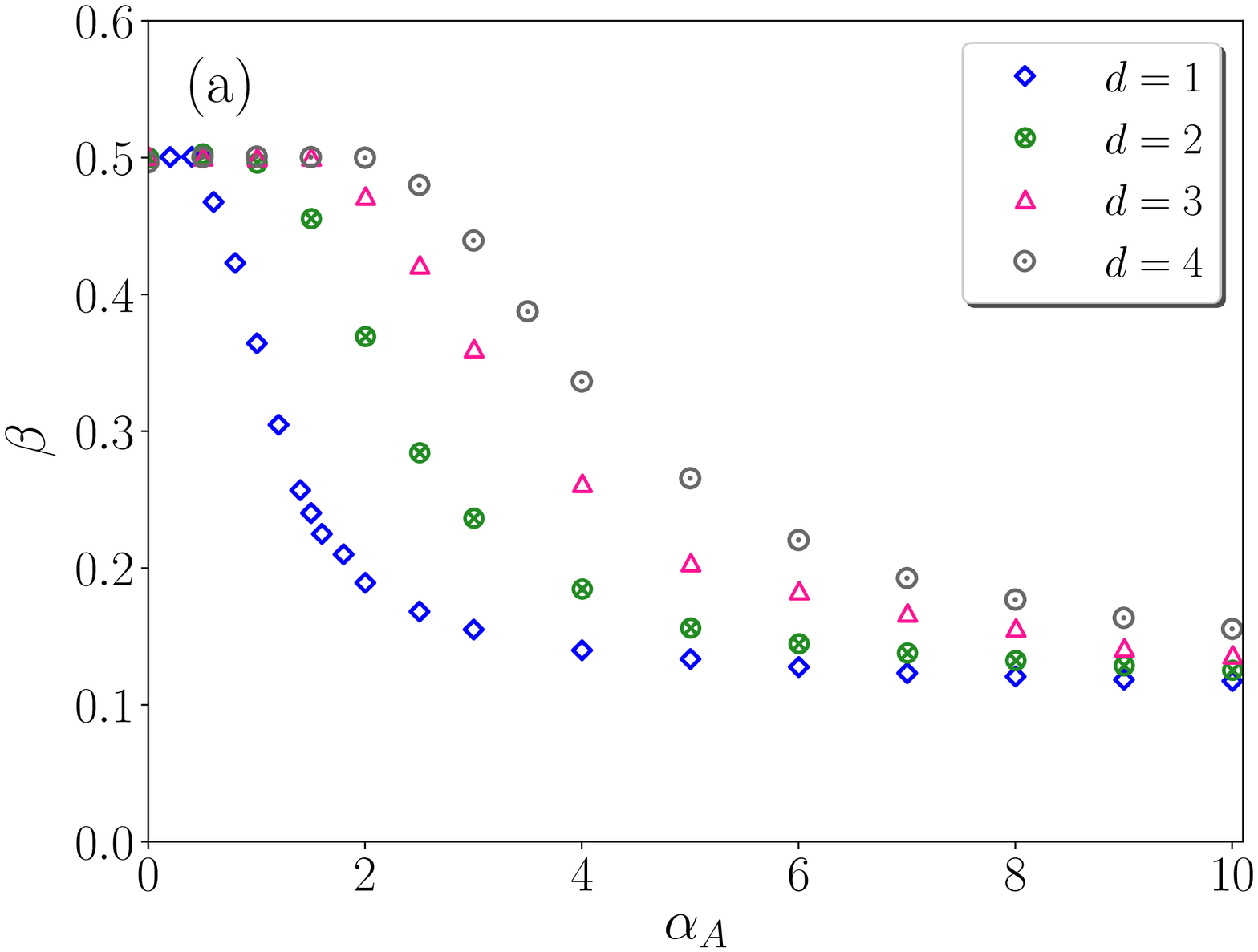} 
\includegraphics[scale=.28]{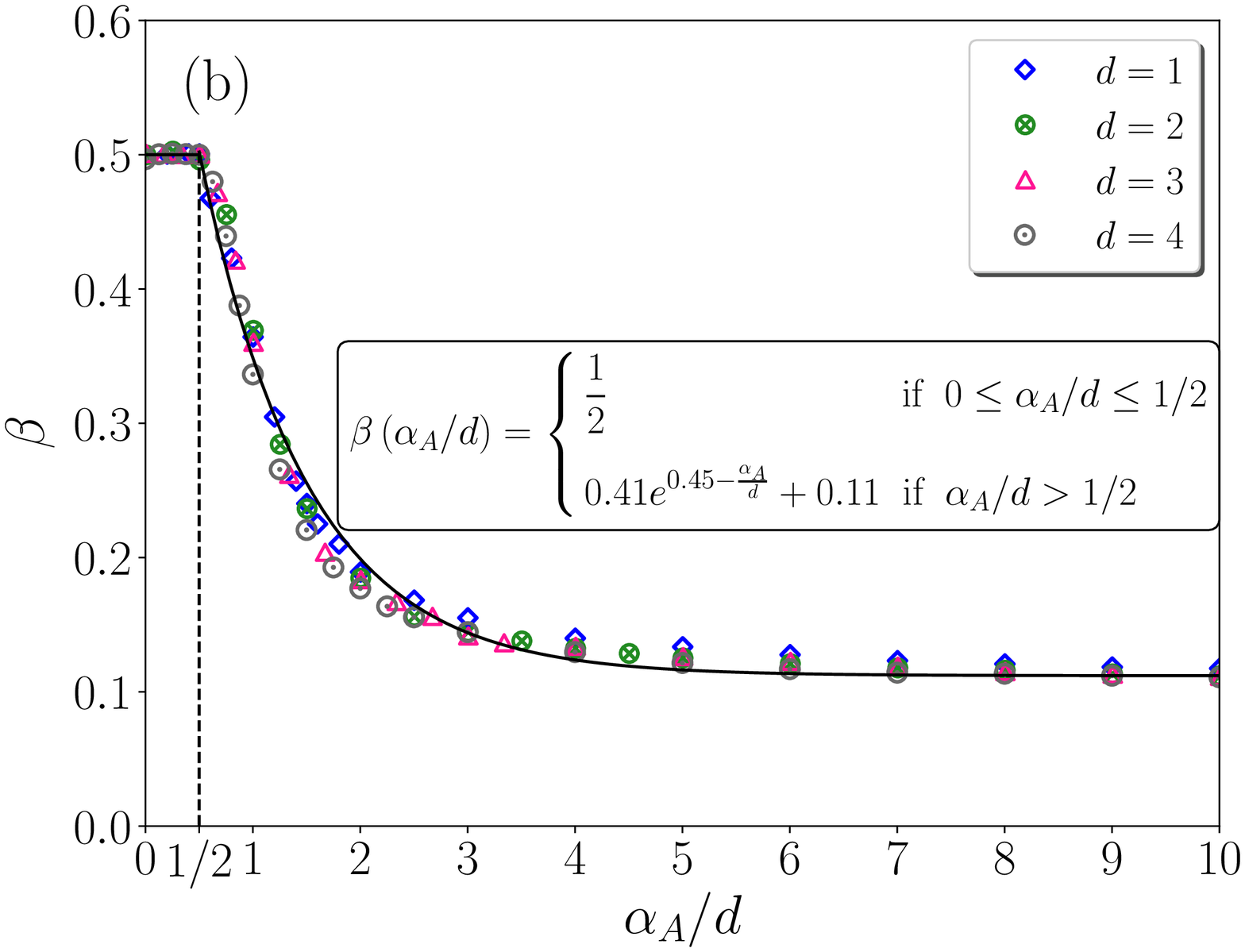}                   
\caption{{\bf(a)} $\beta$ decreases with $\alpha_A$ and increases with $d$. {\bf (b)} As we can see, from the rescaling $\alpha_A \to \alpha_A/d$, all the curves of $\beta$ collapse, and three regions clearly emerge. The first one is from $\alpha_A/d=0$ up to the critical value  $\alpha_A/d=1/2$, $\beta = 1/2$. The second regime, from the critical value $\alpha_A/d= 1/2$ up to the characteristic value  $\alpha_A/d \simeq 5$, $\beta$ decreases nearly exponentiallly. When $\alpha_A/d \gtrsim 5$,  $\beta$ reaches a terminal value $\beta \simeq 0.11$, and the Bolzmannian-like regime is achieved. The simulations have been run for $10^3$ samples and $N = 10^5$.}
\label{beta_alfa_dim}
\end{figure}

We observe that the dynamic exponent $\beta$ is not constant, in discrepancy with its value for the BA model: $\beta$ decreases with $\alpha_A$ and increases with $d$ (see Fig.~\ref{beta_alfa_dim}a). Moreover, we notice that $\beta$ exhibits universal curves with respect to the scaled variable $\alpha_A/d$. When $\alpha_A/d \geq 0$ up to the critical value $\alpha_A/d=1/2$, the system is in the same universality class of the BA model with $\beta = 1/2$ and it is in the non-Boltzmannian very-long-range interactions regime ($q$ is constant above unity). From $\alpha_A/d > 1/2$ on, $\beta < 1/2$ and decreases nearly exponentially with $\alpha_A/d$ down to the value $0.11$ for $\alpha_A/d \approx 5$. For $\alpha_A/d$ above this value up to infinity, $\beta$ remains practically constant, indicating a Boltzmannian-like regime (with $q \simeq 1)$. It cannot be excluded that the terminal value of $\beta$ is achieved only at the limit $\alpha_A/d \to\infty $ (see Fig.~\ref{beta_alfa_dim}b).

\subsection{Average shortest path length}

\begin{figure}[!htb]
\centering
\includegraphics[scale=.29]{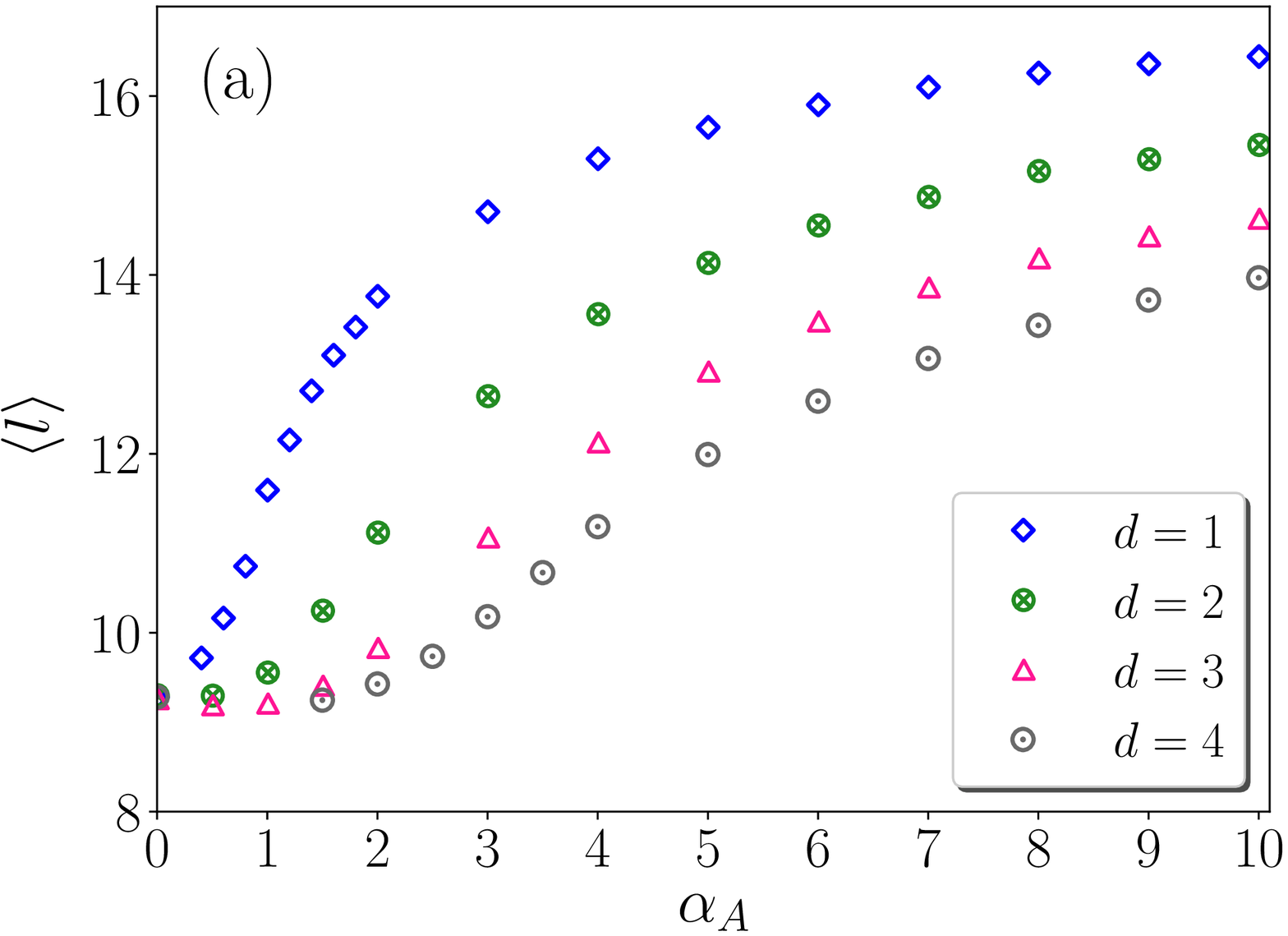} 
\includegraphics[scale=.29]{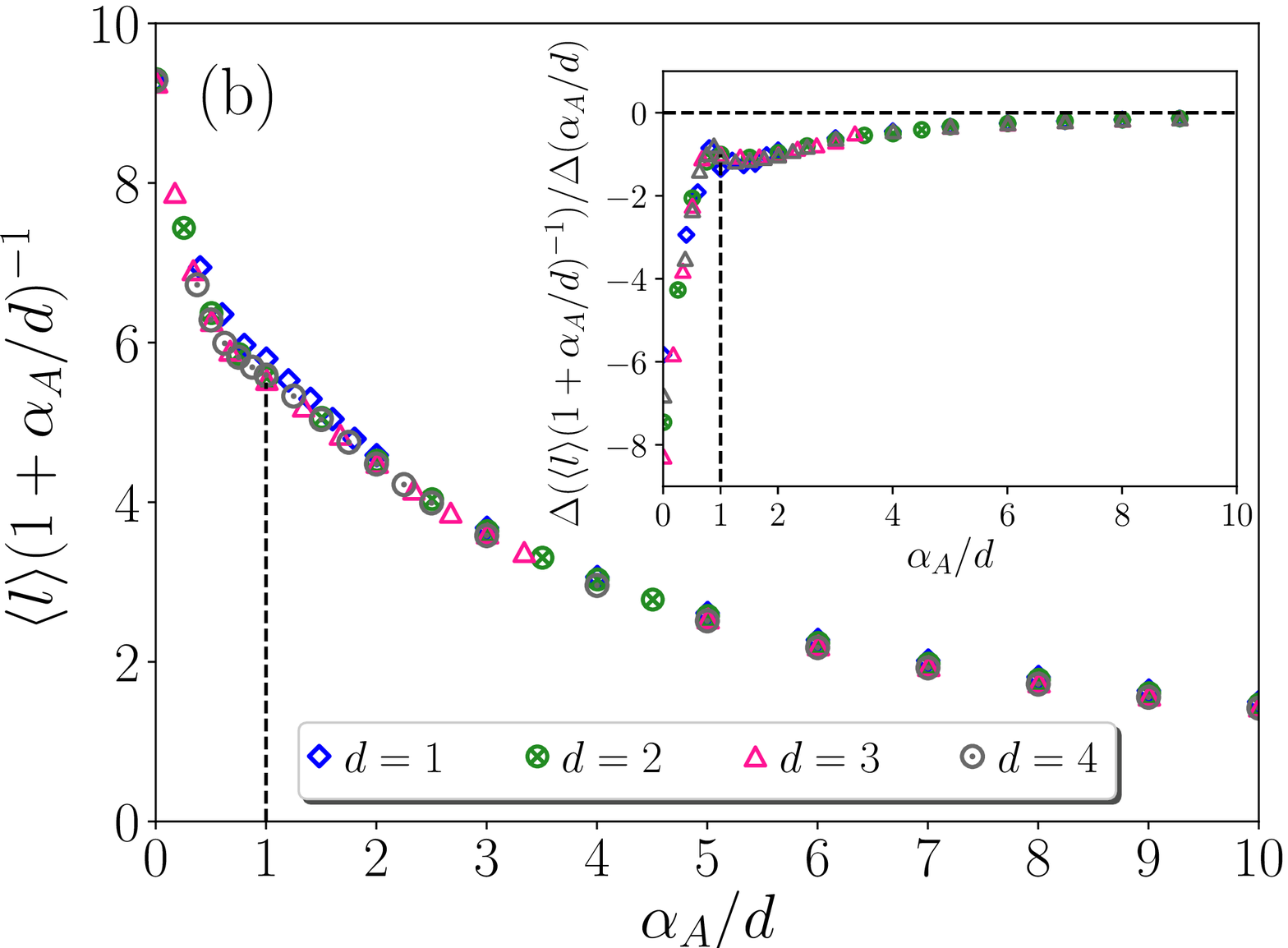}
\caption{Shortest path length. {\bf (a)} We can see that the chemical distance $\langle l \rangle$ increases with $\alpha_A$ and decreases with $d$. {\bf (b)} We can observe that the curves exhibit universality when we re-scale the axis replacing $\alpha_A \to \alpha_A/d$ and $\langle l \rangle \to \langle l \rangle(1+ \alpha_A/d)^{-1}$. In the inset plot we show the derivative collapsed curve in order to see more precisely the existence of the three regimes. The critical value $\alpha_A/d = 1$, show us the end of the non-Boltzmanninan very-long-range interactions. The second regime, the non-Boltzmannian moderate long-range interactions, go to up the characteristic value $\alpha_A/d \simeq 5$. Finally, the third regime, from the characteristic value up to $\alpha_A/d \to \infty$, we see the Boltzmannian-like behavior characterizing short-range interactions. This results are for $N=10^4$ and $10^3$ samples.}  
\label{mc}
\end{figure}

The {\it average shortest path length} is a concept, in network theory, defined as the average number of steps along the shortest paths for all possible pairs of sites of the network. In real networks, a short path makes it easier to transfer information and can reduce  costs. Mathematically, the average shortest path length is defined by
\begin{equation}
\langle l \rangle = \frac{2}{N(N-1)}\sum_{i<j}{d_{ij}},
\end{equation}
where $d_{ij}$ is the shortest path (smaller number of edges) between the sites $i$ and $j$.
We have computed the average shortest path length $\langle l \rangle$ for typical values of $\alpha_A$ and $\forall d$. When $\alpha_A=0$ the results are the same as the BA model where $\langle l \rangle \sim \ln N$ (for $m=1$), independent of the dimension of the system. We have numerically verified that $\langle l \rangle$ depends on ($\alpha_A$,$d$), increasing with $\alpha_A$ and decreasing with $d$ (see Fig.~\ref{mc}a). Remarkably enough,  all the curves can be made to collapse through the scalings $\alpha_A \to \alpha_A/d$ and $\langle l \rangle \to \langle l \rangle (1+ \alpha_A/d)^{-1}$ (see Fig.~\ref{mc}b).  Again, we can see the existence of three regimes. The non-Boltzmannian very-long-range interactions go up to the critical value $\alpha_A/d = 1$, as we can see in the inset plot (Fig.~\ref{mc}b). The non-Boltzmannian moderate long-range interactions go to up to the characteristic value $\alpha_A/d \simeq 5$, as can be seen from the derivative of the collapse curve. And finally, from $\alpha_A/d \gtrsim 5 $ on, the Boltzmannian-like limit is reached.
 
\subsection{Degree distribution entropy}

The computation of the entropy in complex networks is important to verify the heterogeneity and structure of the network~\cite{lewis}. The {\it degree distribution entropy} measures the quantity of randomness  present in the connectivity distribution. In our simulations, it is possible to realize the change of the topology of the network. When $\alpha_A=0$ the network is a scale-free with an asymptotically power law connectivity distribution. As we increase the value of $\alpha_A$ the randomness of the degree distribution also increases. For $\alpha_A\to \infty\, (q \to 1)$ the network is not scale-free anymore since it presents an exponential degree distribution, in agreement with some results available in the literature~\cite{OzikHuntOtt2004}. 

\begin{figure}[!htb]
\centering
\includegraphics[scale=.32]{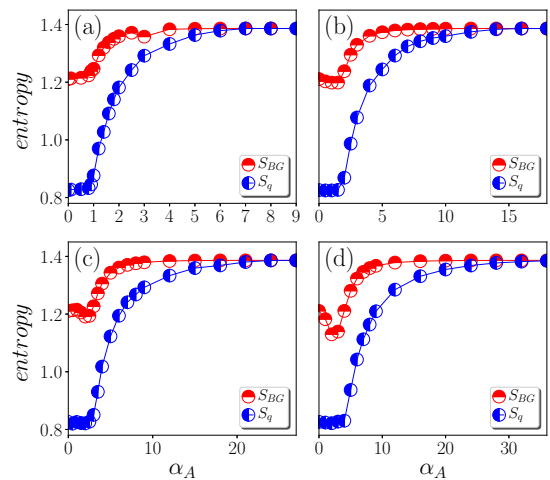} 
\caption{Measure of entropy in complex networks. Comparison between $q-$entropy $(S_q)$ and the standard entropy $(S_{BG})$. The BG entropy was calculated in the network using $S_{BG} = -k\sum_k p_k \ln{p_k}$ (we used $k = 1$), whereas the $q$-entropy was calculated using $S_q = - \sum_k p_k^{q} \ln_q{p_k}$, where $p_k$ is the probability of finding a site with connectivity $k$ and $\sum_k p_k = 1$. In the region of long-range interactions we can see that $S_q$ is very different from $S_{BG}$, exhibiting that $S_q$ is more sensitive for describing this model in this domain. When $\alpha_A \to \infty$ $(q \rightarrow 1)$ both entropies converges to the same asymptotic behaviour. The sublabels refer to {\bf(a)} $d=1$, {\bf(b)} $d = 2$, {\bf(c)} $d = 3$, and {\bf(d)} $d = 4$.}
\label{entropy_compare}
\end{figure}

We have computed the degree distribution entropy $S$ for each value of $\alpha_A$ and $d=1,2,3,4$. We computed the $q$-entropy $(S_q)$, from nonextensive statistical mechanics, and the Boltzmann-Gibbs (BG) $(S_{BG})$ entropy (alternatively referred to as Shannon entropy)  for the same connectivity distributions studied in \cite{BritoSilvaTsallis2016} (see Fig.~\ref{entropy_compare}). The BG entropy was calculated from $S_{BG}= \sum_k p_k \ln (1/p_k)$, where $p_k$ is the probability to find sites with $k$ degree and the sum is over $k=1$ up to $k_{max}$ under the constraint $\sum_k p_k = 1$. Since we have $P(k)\sim {e_q}^{-k/\kappa}$, to each value $\alpha_A/d$, a pair of parameters $(q, \kappa)$ is associated. This enables, in particular, the computation of the $q$-entropy $S_q \equiv  \sum_k p_k \ln_q (1/p_k) $ for the same data, where $\ln_q z \equiv \frac{z^{1-q}-1}{1-q}$ is the inverse of the $q$-exponential function. When $\alpha_A \to \infty$ $(q\to 1)$, both entropies converge to the same asymptotic limit. This result was of course expected since, for $q=1$, the  $q$-entropy recovers the standard entropy $S_{BG}$.

\begin{figure}[!htb]
\centering
\includegraphics[scale=.28]{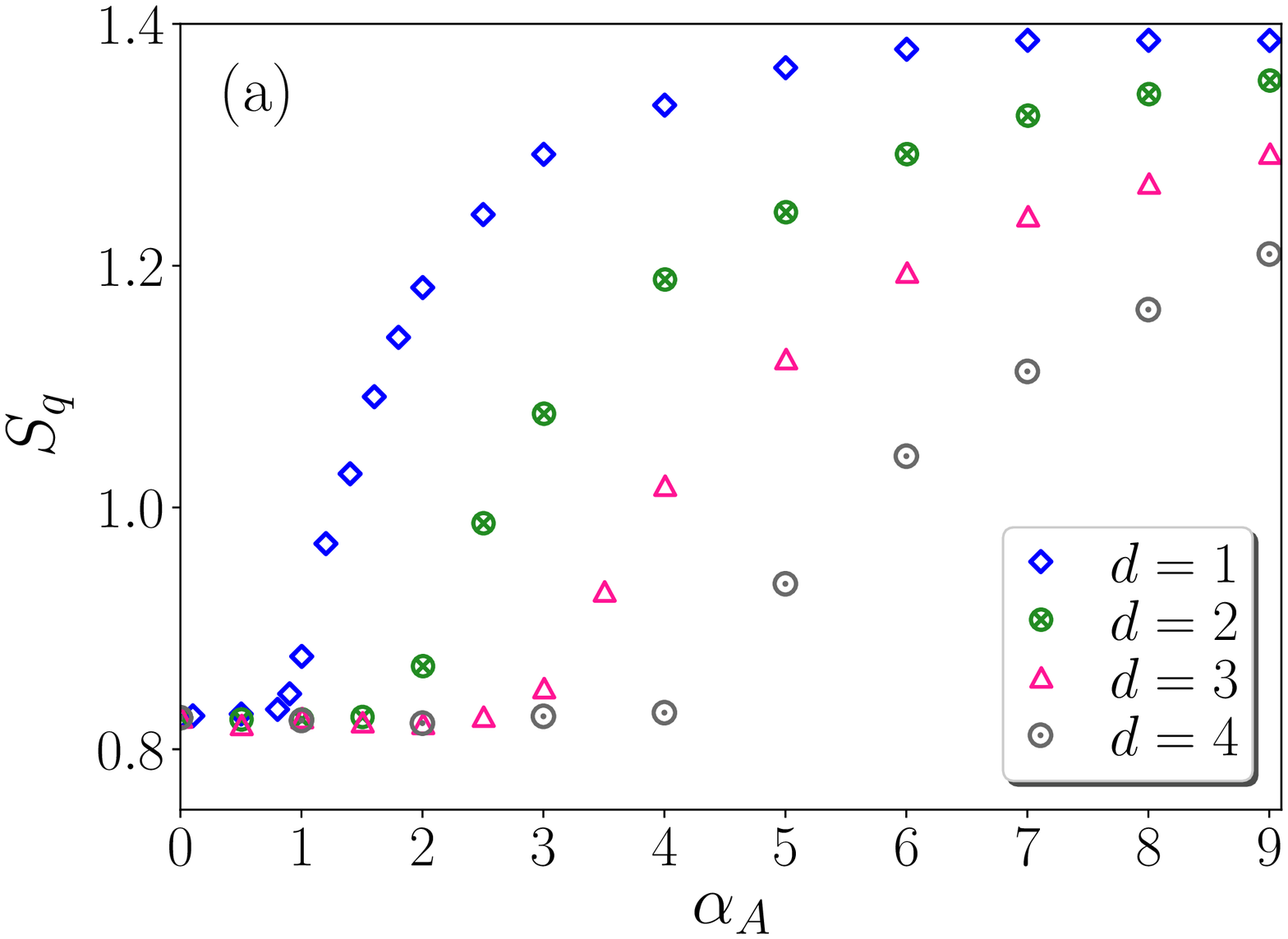} 
\includegraphics[scale=.28]{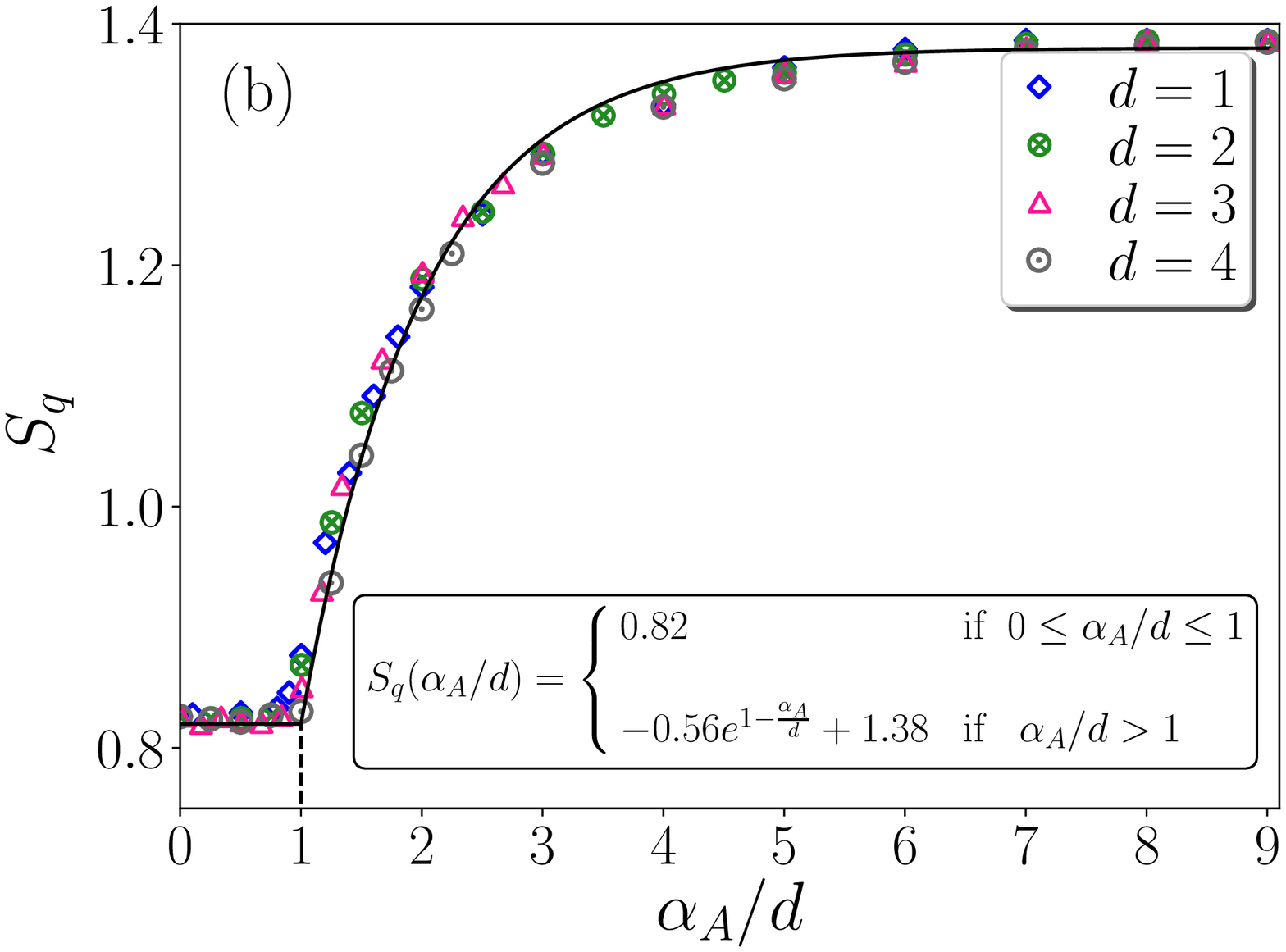}
\caption{Entropy dependence of $\alpha_A$ and $d=1,2,3,4$. {\bf (a)} $S_q$ increases with $\alpha_A$ and decreases with $d$. {\bf (b)} Once again, we obtain the collapse of $S_q$ when rescaling $\alpha_A \to \alpha_A/d$. The entropy $S_q$ has a constant value up to the critical value $\alpha_A=1$ and a nearly exponential behavior emerges up to the characteristic value $\alpha_A \simeq 5$.}
\label{entropy}
\end{figure}

Our results show that there is a region where the two entropies are different. It is known that the BG entropy is not appropriate for systems where long-range interactions are allowed. So, this result provides  evidence that $S_q$ is adequate to describe the interactions in this nonextensive domain. Besides this result, we also studied the dependence of $S_q$ with both $(\alpha_A, d)$ and $\alpha_A/d$. We verified that, although $S_q$ depends on $\alpha_A$ and $d$ separately (see Fig.~\ref{entropy}a), the curves exhibit universal behavior with regard to the scaled variable $\alpha_A/d$ (see Fig.~\ref{entropy}b). Once again, we clearly see the existence of three regimes. In the first one, $S_q$ has a constant value up to the critical value $\alpha_A/d =1$. From that value on, the characteristic value $\alpha_A/d \simeq 5$, $S_q$ increases nearly exponentially and then, from $\alpha_A/d \gtrsim 5$ on, the Boltzmannian-like limit is achieved.

\subsection{Average clustering coefficient}

The {\it average clustering coefficient} is an important measure in the theory of networks and it is associated with how the neighbours of a given node are connected to each other. This coefficient is defined as follows:

\begin{equation}
\langle C \rangle = \frac{1}{N}\sum_i\frac{2n_i}{k_i(k_i - 1)},
\end{equation}
where $k_i$ is the degree of the site $i$, $n_i$ is the number of connections between the neighbours of the site $i$ and $k_i(k_1 - 1)/2$ is the total number of possible links between them. 

\begin{figure}[!htb]
\centering
\includegraphics[scale=.28]{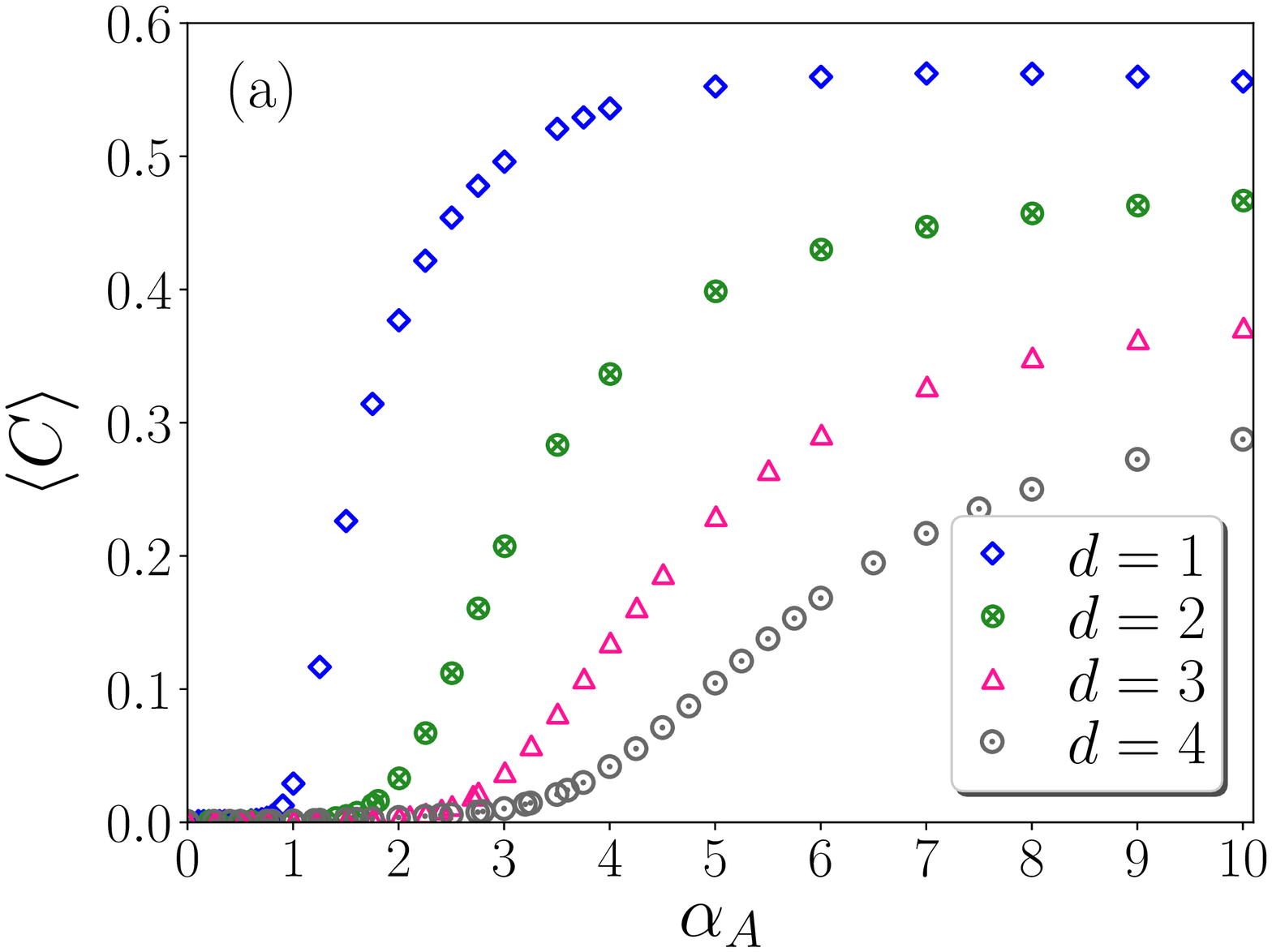} 
\includegraphics[scale=.28]{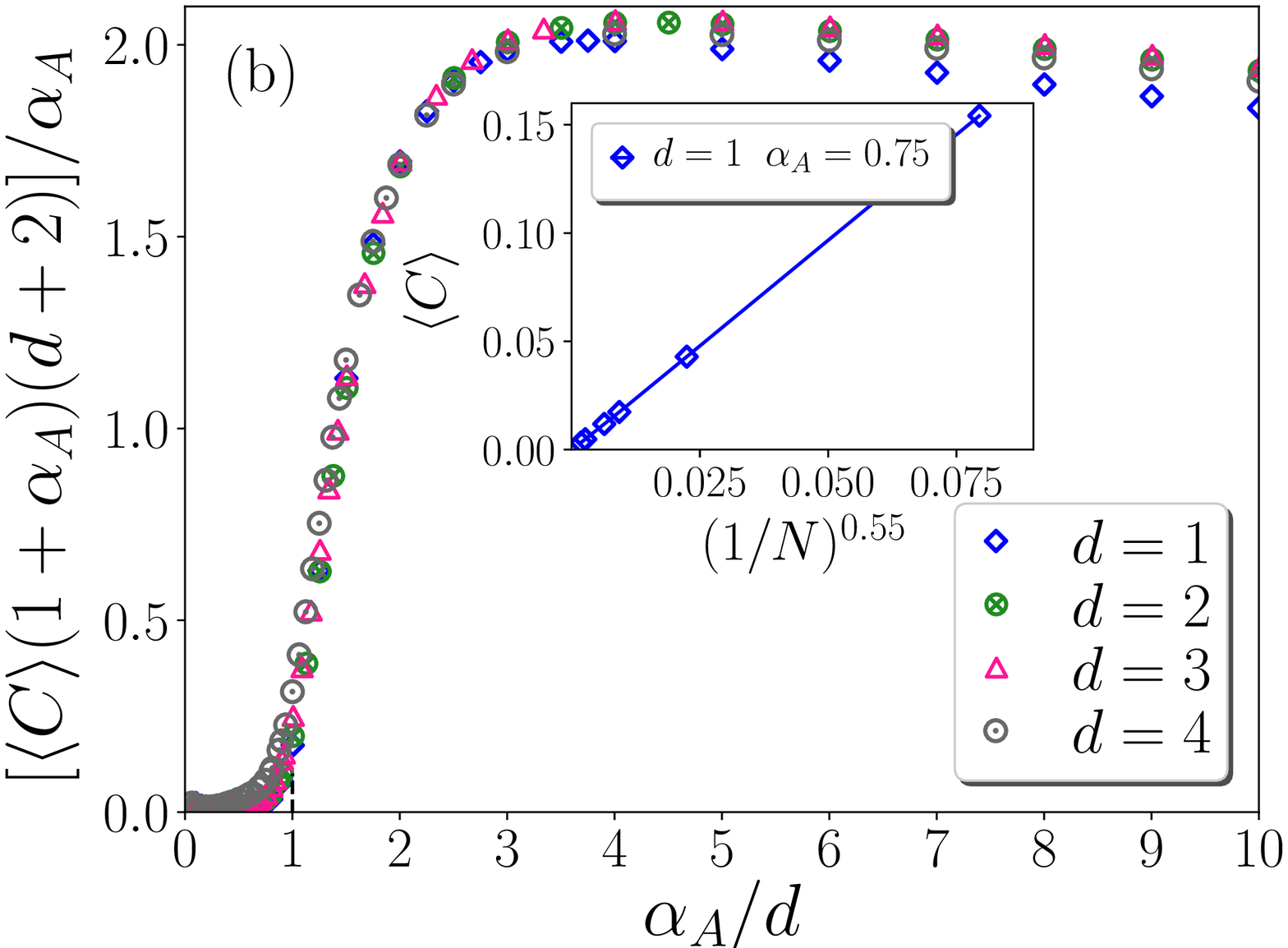}
\caption{Clustering coefficient
for typical values of $\alpha_A$ and $d=1,2,3,4$. {\bf (a)} $\langle C \rangle$ increases with $\alpha_A$ but it decreases with $d$.  {\bf (b)} All curves collapse with the rescaling $\langle C \rangle \to \langle C \rangle [(\alpha_A +1)(d +2 )]/\alpha_A$ and $\alpha_A \to \alpha_A/d$. In the thermodynamical limit $ \langle C \rangle \to 0$  from $\alpha_A = 0$ up to the critical value $\alpha_A/d = 1$. In the inset plot we show an example for $d=1$ and $\alpha_A=0.75$. From the characteristic value $\alpha_A/d \simeq 5$ on we reach the Boltzmannian-like regime.}
\label{clustering}
\end{figure}

\begin{figure}[!htb]
\centering
\includegraphics[scale=.40]{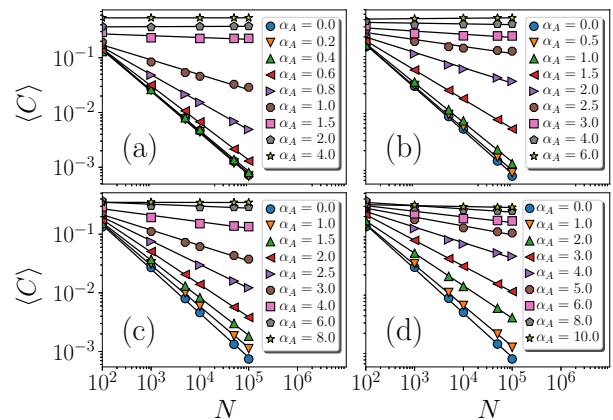} 
\caption{Clustering coefficient as a function of $N$ for {\bf (a)} $d=1$, {\bf (b)} $d = 2$, {\bf (c)} $d = 3$, and {\bf (d)} $d = 4$. $\langle C \rangle$ decreases with $N$, but increases with $\alpha_A$. Interestingly, from $\alpha_A \sim 2d$ on, $\langle C \rangle$ does not change any more with $N$.}
\label{C_N_1d2d3d4d}
\end{figure}

\begin{figure}[!htb]
\centering
\includegraphics[scale=.28]{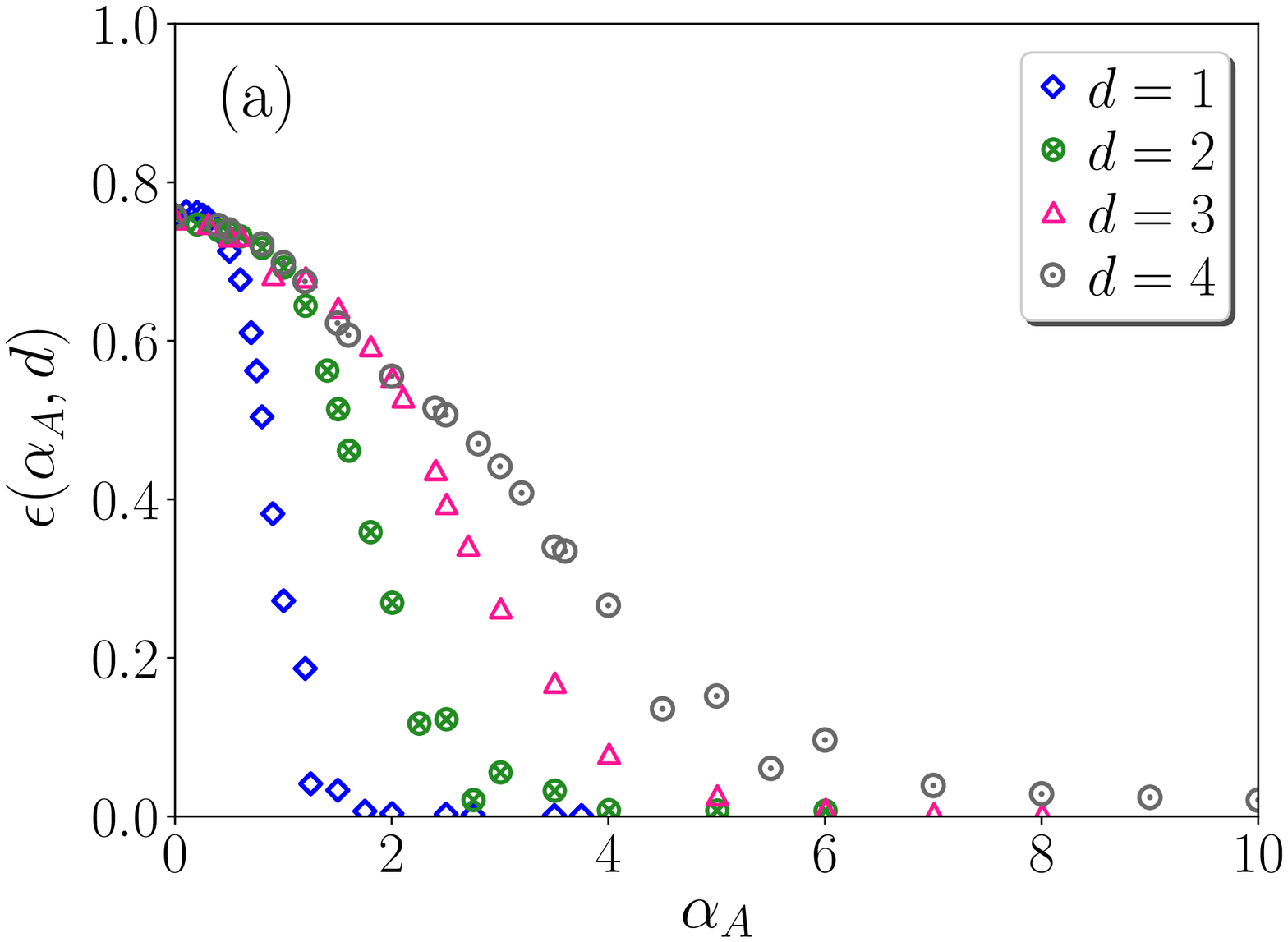} 
\includegraphics[scale=.28]{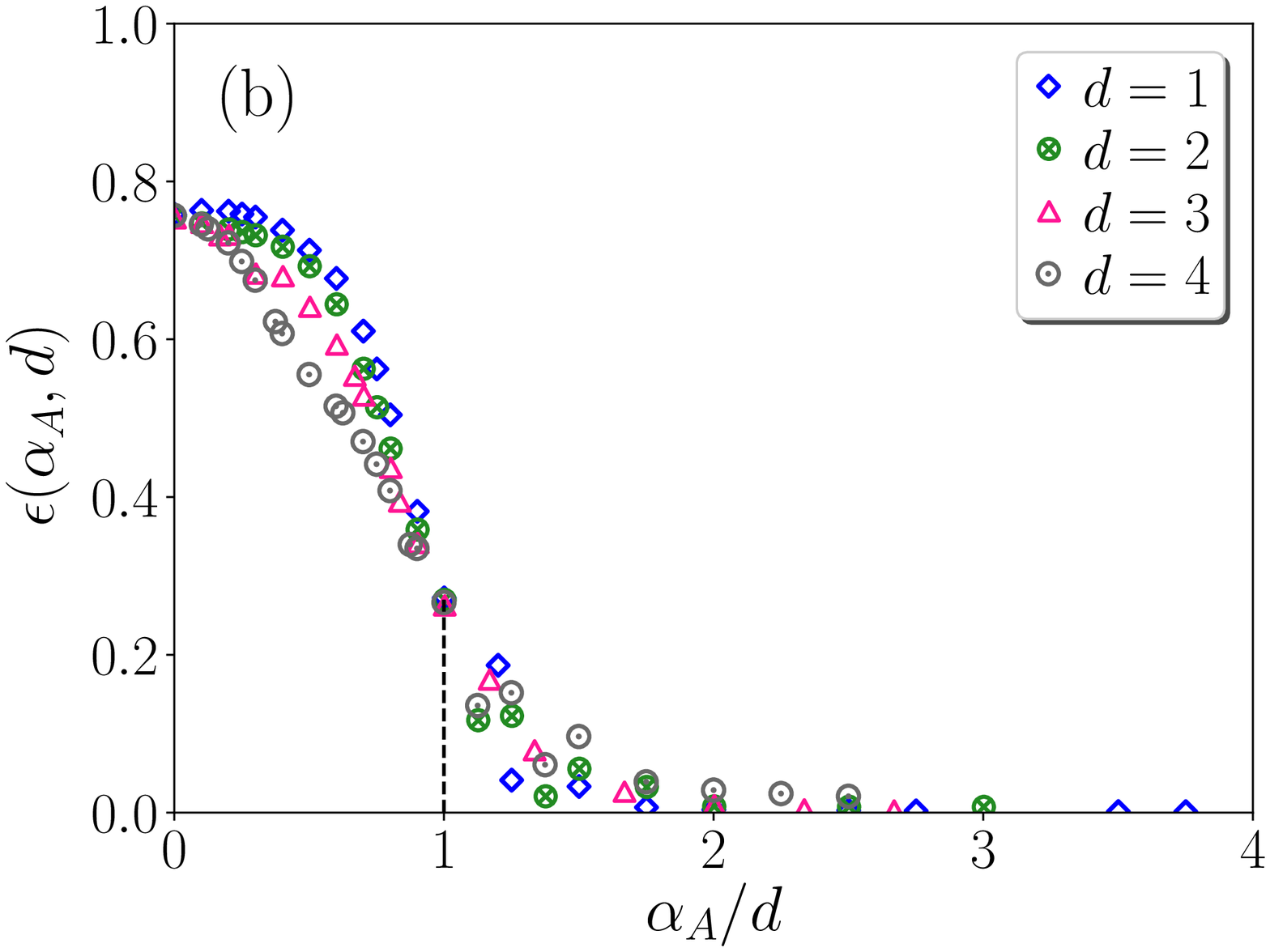} 
\caption{Analysis of the exponent $\epsilon(\alpha_A,d)$. {\bf (a)} This graph reminds us the behavior of the maximum Lyapunov exponent \cite{BagchiTsallis2016}. The exponent $\epsilon(\alpha_A,d)$ decreases with $\alpha_A$ and does so faster for smaller $d$. {\bf (b)} Although we did not get collapse for these curves through the rescaling $\alpha_A \to \alpha_A/d$, we can clearly see that all curves appear to perfectly intersect at $\alpha_A/d = 1$. This result strongly indicates a change of regime, with something special occurring at the value $\alpha_A/d = 1$. When $\alpha_A/d \gtrsim 2$ we observe that $\epsilon \to 0$ agrees with the result showed in Fig.~\ref{C_N_1d2d3d4d}.}
\label{epsilon}
\end{figure}

In order to compute it, we run our network model for $m=2$ (because $\langle C \rangle=0$ when $m=1$) and analyzed how $\langle C \rangle$ changes with both $(\alpha_A, d)$ and $\alpha_A/d$. We see that  $\langle C \rangle$ increases with $\alpha_A$ and decreases with $d$ (see Fig.~\ref{clustering}a). The larger $\alpha_A$ the more aggregated the network is. In the standard Barab\'asi-Albert model $(\alpha_A = 0)$, the clustering coefficient is influenced by the size $N$ of  the network, such that $\langle C \rangle$ can be numerically approximated by  $\langle C \rangle \sim N^{-0.75}$ (in later works, Barab\'asi analytically claimed that $\langle C \rangle \sim (\log N)^2/N$; for further details see~\cite{NetworkScienceBA}). From this behaviour we can see that, in the thermodynamical limit $(N \to \infty)$, $\langle C \rangle \to 0$. So, we have numerically verified that when $N \to \infty$ $\langle C \rangle \to 0$ not only for $\alpha_A/d = 0$, but for $0 \leq \alpha_A/d \leq 1$ (see the inset plot in Fig.~\ref{clustering}b). We also analyzed how the clustering coefficient changes with $N$ and we found that $\langle C \rangle \sim N^{-\epsilon(\alpha_A,d)}$. This power-law form was in fact expected since it agrees with the numerical result previously found for the particular case $\alpha_A = 0$. However, surprisingly enough, when $\alpha_A \gtrsim 2d$ the clustering coefficient does not change with $N$ anymore (see Fig.~\ref{C_N_1d2d3d4d}). Analyzing how $\epsilon(\alpha_A,d)$ changes with both $(\alpha_A,d)$ and $\alpha_A/d$, we see that this exponent decreases with $\alpha_A$, but increases with $d$ (see Fig.~\ref{epsilon}a). Although we did not get collapse for these curves, by rescaling $\alpha_A \to \alpha_A/d$ we clearly can see that all curves perfectly intersect in $\alpha_A/d = 1$, strongly indicating a change of regime (see Fig.~\ref{epsilon}b). The results found for $\epsilon(\alpha_A,d)$ are somewhat reminiscent of the $\kappa(\alpha_A,d)$ exponent associated with the maximal Lyapunov exponent for the generalized Fermi-Pasta-Ulam (FPU) model \cite{BagchiTsallis2016}.

\section*{Conclusion}

Our present results reveal an intriguing ubiquity of the variable $\alpha_A/d$ for the class of networks focused on here, where both topological and metric aspects exist. Surprisingly, the use of this variable indeed provides collapses or quasi collapses for all the properties studied here. Another interesting point is the existence of three, and not only two, regimes. The first one is a non-Boltzmannian regime characterized by very-long-range interactions and it goes from $\alpha_A/d=0$ up to the critical value $\alpha_A/d =1$, except for $\beta$ whose critical value turns out to be $\alpha_A/d =1/2$, curiously enough.  The second one, from the critical value up to the characteristic value $\alpha_A/d \simeq 5$, is still non-Boltzmannian and corresponds to moderate long-range interactions. The third and last regime, from $\alpha_A/d \simeq 5$ on, is Boltzmannian-like and is characterized by short-range interactions. The existence of the intermediate regime has also been observed in classical many-body Hamiltonians, namely the $\alpha$-generalized XY \cite{CirtoAssisTsallis2014,CirtoRodriguezNobreTsallis2018} and Heisenberg ~\cite{CirtoLimaNobre2015} rotator models as well as the Fermi-Pasta-Ulam model ~\cite{christo1,christo2,BagchiTsallis2016,BagchiTsallis2017,bagchipassage}.

The present work neatly illustrates a fact which is not always obvious to the community working with complex networks, more precisely those exhibiting asymptotic scale-free behavior.  Such networks frequently belong to the realm of applicability of nonextensive statistical mechanics based on nonadditive entropies, and to its superstatistical extensions \cite{BeckCohen}. Such connections between thermal and geometrical systems are by no means rare in statistical mechanics since the pioneering and enlightening Kasteleyn and Fortuin theorem ~\cite{KasteleynFortuin1969}. In the present scenario, such connection can be naturally understood if we associate half of each two-body interaction energy between any two sites of the Hamiltonian to each of the connected sites, thus generating, for each node, a degree (number of links) in the sense of networks. Through this perspective, it is no surprise that the degree distribution corresponds to the $q$-exponential function which generalizes the Boltzmann-Gibbs weight within thermal statistics. The fact that in both of these geometrical and thermal systems, the scaled variable $\alpha_A/d$ plays a preponderant role becomes essentially one and the same feature. Mathematically-based contributions along such lines would be more than welcome. Last but not least, it would surely be interesting to understand how come the critical point of the $\beta$ exponent differs from the all the critical points that we studied here. This is somewhat reminiscent of the two-dimensional XY ferromagnetic model with short-range interactions for which nearly all properties exhibit a singularity at the {\it positive} temperature of Kosterlitz and Thouless~\cite{KosterlitzThouless1973,Tsallis1976}, whereas the order parameter critical point occurs at {\it zero} temperature.


\section*{Acknowledgments}
We gratefully acknowledge partial financial support from CAPES, CNPq, Funpec, Faperj (Brazilian agencies) and the Brazilian ministries MEC and MCTIC. We thank the High Performance Computing Center at UFRN for providing the computational facilities to run the simulations.


\newpage
\begin{thebibliography}{9}


\bibitem{Barabasi2011} A. L. Barab\'asi,  N. Gulbahce, and J. Loscalzo,  Nature Reviews Genetics {\bf12}(1), 56-68 (2011).

\bibitem{Boguna2014} M. Bogu\~n\'a, M. Kitsak, and D. Krioukov, New Journal of Physics  {\bf16}(9), 093031 (2014).

\bibitem{Perseguers2010} S. Perseguers, M. Lewenstein, A. AcÃ­n, and J. I. Cirac, Nature Physics  {\bf6}(7), 539-543 (2010).

\bibitem{Ferreira2018} G. D. Ferreira, G. M. Viswanathan, L. R. da Silva and H. J. Herrman,  Physica A {\bf499}, 198-207 (2018).

\bibitem{SoaresTsallisMarizSilva2005} D. J. B. Soares, C. Tsallis, A. M. Mariz, and L. R. da Silva, EPL {\bf 70}(1), (2005).

\bibitem{ThurnerTsallis2005} S. Thurner and C. Tsallis, EPL {\bf 72}(2), 197 (2005).

\bibitem{Thurner2005} S. Thurner, Europhysics News {\bf 36}(6), 218-220 (2005).

\bibitem{Andradeetal2005} J. S.  Andrade Jr., H. J. Herrmann,  R. F. Andrade, and L. R. da Silva, Phys. Rev. Lett. {\bf 94}(1), 018702 (2005).

\bibitem{Lindetal2007} P. G. Lind, L. R. da Silva, J. S. Jr. Andrade,  H. J. Herrmann, Phys. Rev. E {\bf 76}(3), 036117 (2007).

\bibitem{Mendesetal2012} G. A. Mendes, L. R. da Silva, and H. J. Herrmann, Physica A {\bf 391}(1), 362-370 (2012).

\bibitem{Almeidaetal2013} M. L. Almeida, G. A. Mendes, G. M. Viswanathan,  and L. R. da Silva, European Phys. J. B {\bf 86}(2), 1-6 (2013).

\bibitem{Luciano} A. Macedo-Filho, D. A. Moreira, R. Silva, and L. R. da Silva, Phys. Lett. A {\bf 377}(12), 842-846 (2013).

\bibitem{BritoSilvaTsallis2016} S. Brito, L. R. da Silva and C. Tsallis, Scientific Reports {\bf 6}, 27992 (2016).

\bibitem{NunesBritoSilvaTsallis2017}  T. C. Nunes, S. Brito, L. R. da Silva and C. Tsallis, J. Stat. Mech {\bf9}, 093402 (2017).

\bibitem{TirnakliBorges2016}  U. Tirnakli and E.P. Borges, Scientific Reports {\bf 6}, 23644 (2016).

\bibitem{CirtoAssisTsallis2014} L. J. L. Cirto, V. R. V. Assis and C. Tsallis, Physica A {\bf 393},  286-296 (2014).   

\bibitem{CirtoRodriguezNobreTsallis2018}L. J. L. Cirto, A. Rodriguez, F. D. Nobre and C. Tsallis, EPL {\bf 123}, 30003 (2018).

\bibitem{CirtoLimaNobre2015}L. J. L. Cirto, L. S. Lima and F. D. Nobre, J. Stat. Mech P04012 (2015).

\bibitem{christo1}H. Christodoulidi, C. Tsallis and T. Bountis, EPL {\bf 108}, 40006 (2014).

\bibitem{christo2}H. Christodoulidi, T. Bountis, C. Tsallis, and L. Drossos, J. Stat. Mech, 123206 (2016).

\bibitem{note1} By the {\it characteristic value} we mean a value above which q is noticeably close to one, whatever strictly speaking the numerical evidence that we have, suggest that the Botzmannian limit is only achieved for $\alpha_A/d \to \infty$. 

\bibitem{Tsallis1988} C. Tsallis, J. Stat. Phys. {\bf 52}, 479 (1988).

\bibitem{Tsallis2009} C. Tsallis. Introduction to Nonextensive Statistical Mechanics - Approaching a Complex World. Springer, New York, 2009. 

\bibitem{TsallisCirto2013}C. Tsallis and L. J. L. Cirto, Eur. Phys. J. C {\bf 73}, 2487 (2013).

\bibitem{BarabasiAlbert1999} A. L. Barab\'asi and R. Albert, 
Science {\bf 286}, 509 (1999).

\bibitem{lewis}T. G. Lewis. Network science: Theory and applications. John Wiley \& Sons, (2011).
 
\bibitem{OzikHuntOtt2004} J. Ozik, B. R. Hunt and E. Ott,  
Phys. Revi. E, {\bf 69} (2), 026108  (2004).

\bibitem{NetworkScienceBA} A. L. Barab\'asi. Network Science (Cambridge University Press, 2016).
 
\bibitem{BagchiTsallis2016} D. Bagchi and C. Tsallis, 
Phys. Rev. E {\bf 93}(5), 062213 (2016).

\bibitem{BagchiTsallis2017} D. Bagchi and C. Tsallis, 
Phys. Lett. A  {\bf 381}, 1123-1128 (2017).

\bibitem{bagchipassage} D. Bagchi and C. Tsallis, 
Physica A {\bf 491}, 869-873 (2018).

\bibitem{BeckCohen} C. Beck and E. G. D. Cohen, 
Physica A {\bf 322}, 267 (2003).

\bibitem{KasteleynFortuin1969} P. Kasteleyn and C. Fortuin, J. Phys. Sot. Jap. (suppl.) {\bf 26},  1 (1969). 

\bibitem{KosterlitzThouless1973} J. M. Kosterlitz and D. J. Thouless,
Phys.: Condensed Matter {\bf 6}, 1 (1973).

\bibitem{Tsallis1976} C. Tsallis,
Il Nuovo Cimento B (1971-1996) {\bf 34} (2), 411-435 (1976).

\end{thebibliography}
\end{document}